\documentclass[10pt, twocolumn]{article}
\usepackage[english]{babel}
\usepackage{amsmath, amssymb, graphicx, cite}
\usepackage{bbold, bm}
\usepackage{subfigure}

\def\be{\begin{eqnarray}}
\def\ee{\end{eqnarray}}

\def\rd{\mathrm{d}}
\newcommand{\SU}{\text{SU}}
\newcommand{\R}{\mathbb{R}}
\newcommand{\su}{\mathfrak{su}}

\begin{document}

\title{Point particles in 2+1 dimensions: toward a semiclassical loop gravity formulation}
\author{Jonathan Ziprick \\
  \multicolumn{1}{p{.7\textwidth}}{\centering\emph{Perimeter Institute and University of Waterloo, \\Waterloo, Ontario, Canada}}}

\maketitle

\begin{abstract}
We study point particles in 2+1 dimensional first order gravity using a triangulation to fix the connection and frame-field. The Hamiltonian is reduced to a boundary term which yields the total mass. The triangulation is dynamical with non-trivial transitions occurring when a particle meets an edge. This framework facilitates a description in terms of the loop gravity phase space.
\end{abstract}


\section{Introduction}
Three dimensional gravity has often been used as a toy model for the four dimensional theory. Here we study point particles in 2+1 dimensions using spatial geometries analogous to those in \cite{FGZ}. These geometries are isomorphic to a gauge-reduced holonomy-flux phase space which descends from the $\hbar \rightarrow 0$ limit of a loop quantum gravity Hilbert space.

We consider a Hamiltonian written in terms of frame-field and connection variables. The spacetime signature is taken to be Euclidean since the relevant gauge group is then $\SU(2)$, as in 3+1 dimensions with Lorentzian signature. Our first step is to specify the variables over an entire spacelike slice using a triangulation where particles sit on the vertices. After gauge fixing, the Hamiltonian is reduced to a boundary term which equals the total particle mass. In this gauge the triangulation evolves according to particle momenta and undergoes discrete changes when a particle meets an edge, collapsing one of the triangles.

\section{Hamiltonian formulation}
Spacetime is taken to be $M=\R \times \Sigma$ where $\Sigma$ is a spacelike surface homeomorphic to a disc. The first order formalism of general relativity parameterizes the gravitational field in terms of a connection $\mathbf{A}$ and a frame-field $\mathbf{e}$, both being one-forms on $\Sigma$ taking values in the $\su(2)$ algebra. We use $\su(2)$ basis elements $\tau^i$ (for $i=0,1,2$) which are given by $-i/2$ times the Pauli matrices. Our notation is such that elements of $\su(2)$ are written as $\mathbf{A} \equiv A^i \tau^i$, for example, and all internal indices are written as superscripts.

In units where $8 \pi G = c = 1$ a Hamiltonian for pure gravity is given by \cite{FL}:
\be
\label{hamiltonian}
H &=& -\int_\Sigma \left( N^i F^i + \lambda^i G^i \right) + \int_{\partial \Sigma} N^i A^i, \nonumber \\
F^i &=& \rd A^i + \frac{1}{2}\epsilon^{ijk} A^j \wedge A^k , \\
G^i &=& \rd e^i + \epsilon^{ijk} A^j \wedge e^k . \nonumber
\ee
$N^i$ and $\lambda^i$ are Lagrange multipliers for the flatness and Gauss constraints respectively. We normalize the time coordinate by choosing $|N|^2 = 1$. The boundary term and the condition $\left. \lambda^i \right|_{\partial \Sigma}=0$ ensure that the variational principal is well-defined and the constraints are first class.
The Poisson algebra is:
\be
\left\{A_a^i(x), e_b^j(y) \right\} &=& \epsilon_{ab} \delta^{ij} \delta^2(x-y) ,
\ee
where $a = 1,2$ labels the space coordinates.

It is well-known that in 2+1 dimensions point particles represent conical singularities \cite{DJT}. We introduce particles by replacing the flatness constraint with $F^i - \sum_I \delta^2(x-q_\pi) p_\pi^i \rd x^2$ so that each particle $\pi$ gives a contribution to the curvature proportional to its momentum $\mathbf{p}_\pi$ at the location $x=q_\pi$ \cite{FL}. In this treatment the particles are defined by the fields $(\mathbf{A},\mathbf{e})$ and do not have independent parameters of their own. We consider the total mass to be less than $2\pi$ so that $\Sigma$ is open \cite{DJT}.

\section{Gauge fixing}

In order to solve the constraints and reduce the Hamiltonian,  we triangulate $\Sigma$ according to the particle locations so that we may give a piecewise definition of $(\mathbf{A},\mathbf{e})$ on $\Sigma \setminus \left\{ q_\pi \right\}$. See \cite{tri} for other approaches along these lines. For clarity, we take the boundary $\partial \Sigma$ to be triangular; the generalization to arbitrary polygons follows simply. We take each particle position to be an internal vertex and connect all vertices with straight edges to obtain a triangulation. There is ambiguity in this procedure, but the number of triangles $T$ is fixed by the number of particles $\Pi$ to be $T = 2(\Pi+1)-1$ via the Euler characteristic for planar graphs. Different triangulations generally lead to different gauge choices.

Within each triangle $\Delta$, we must choose  $(\mathbf{A}_\Delta,\mathbf{e}_\Delta)$ such that $G^i = F^i = 0$. The general solution \cite{FGZ} is given by an $\SU(2)$ group element $a_\Delta$ and a closed one-form ${\bm \chi}_\Delta$:
\be
\label{solutions}
\mathbf{A}_\Delta = a_\Delta \rd a_\Delta^{-1}, \hspace{0.5in} \mathbf{e}_\Delta = a_\Delta {\bm \chi}_\Delta a_\Delta^{-1}.
\ee
Consider two adjacent triangles labeled $1$ and $2$. We ensure that the fields are continuous across the shared edge $e$ by requiring that there exists an $h_e \in \SU(2)$ on the edge such that:
\be
\label{gluing}
\left. a_2 \right|_e = \left.a_1 h_e \right|_e, \hspace{0.5in} \left. {\bm \chi}_2\right|_e = \left. h_e^{-1}  {\bm \chi}_1 h_e \right|_e.
\ee

Now, if we integrate the curvature over a region containing a single vertex $\pi$ we must obtain $\int F^i = p_\pi^i$. In order to specify the group elements $a_\Delta$ we subdivide each triangle into three regions defined by edges joining the centroid to the vertices.
Consider one such region $\Delta_r$ where points $(0,\pm y_0)$ are the endpoints of an edge, and $(x_0,y_1)$ is the centroid of the triangle. We define:
\be
u =  \tanh^{-1} (v_{+}) + \tanh^{-1} (v_{-}),
\ee
where $v_{\pm} = \frac{x}{x_0} \pm \left( \frac{y}{y_0}-\frac{x y_1}{x_0 y_0} \right)$. Note that the line $u=0$ coincides with the edge, and $u=\infty$ consists of lines between the centroid of the triangle and the endpoints of the edge.
We set the group element in a region equal to:
\be
\label{gena}
a_{\Delta_r}(u)= \exp \left( \mathbf{P} \int_0^u f(\tilde{u}) \rd \tilde{u} \right),
\ee
where we have introduced a bump function $f(u)$ which satisfies $\int_0^\infty f(u) \rd u = 1$. For a given $u$, this group element defines a rotation about a direction $\mathbf{P}/|P|$ by an angle proportional to $|P|$.

With these ingredients it is now possible to choose a gauge, i.e. to specify the frame-field and connection on $\Sigma \setminus \left\{ q_\pi \right\}$. Given a set of particles $\left\{q_\pi, p_\pi\right\}$, one first defines a boundary and introduces a triangulation. Then we arbitrarily choose a triangle (labeled 1) and make a choice for $(a_1, {\bm \chi}_1)$. Moving to a neighbouring triangle, we specify $a_2$ here and use (\ref{gluing}) to find ${\bm \chi}_2$. We continue in this manner until $(a_\Delta, {\bm \chi}_\Delta)$ (and thereby $(\mathbf{A}_\Delta, \mathbf{e}_\Delta))$ are defined in each triangle, taking care to ensure the connection yields proper curvature at each vertex by choosing the parameter $\mathbf{P}$ in each $a_{\Delta_r}$ to be consistent with particle momenta. The aforementioned relationship between the number of triangles and the number of particles ensures that this procedure will work for arbitrarily many particles. See \cite{Z} for a more detailed description of this process.


Preserving the gauge dynamically leads to a condition on the Lagrange multipliers:
\be
\label{condition}
\rd_A N^i + \epsilon^{ijk}e^j\lambda^k = 0 .
\ee
This fixes three of the six degrees of freedom in $N^i$ and $\lambda^i$. We eliminate the remaining ambiguity by choosing $\rd_A N^i =0$ and setting $\mathbf{N}(q_\pi) = \mathbf{p_\pi}/m$. This choice ensures that particles move along with the triangulation and remain at the vertices.

\section{Dynamics and observables}

After specifying $(\mathbf{A},\mathbf{e})$, the Hamiltonian is reduced to a boundary term which evaluates to:
\be
H &=& \int_{\partial \Sigma} N^i A^i = \int_\Sigma N^i \rd_A A^i \nonumber \\
&=& \sum_\pi N^i(q_\pi) p_\pi^i = \sum_\pi m_\pi,
\ee
where we used Stokes' theorem, the conditions on $\mathbf{N}$, and the flatness constraint.

\begin{figure}[htb!]
\begin{center}
\includegraphics[width=0.85\linewidth]{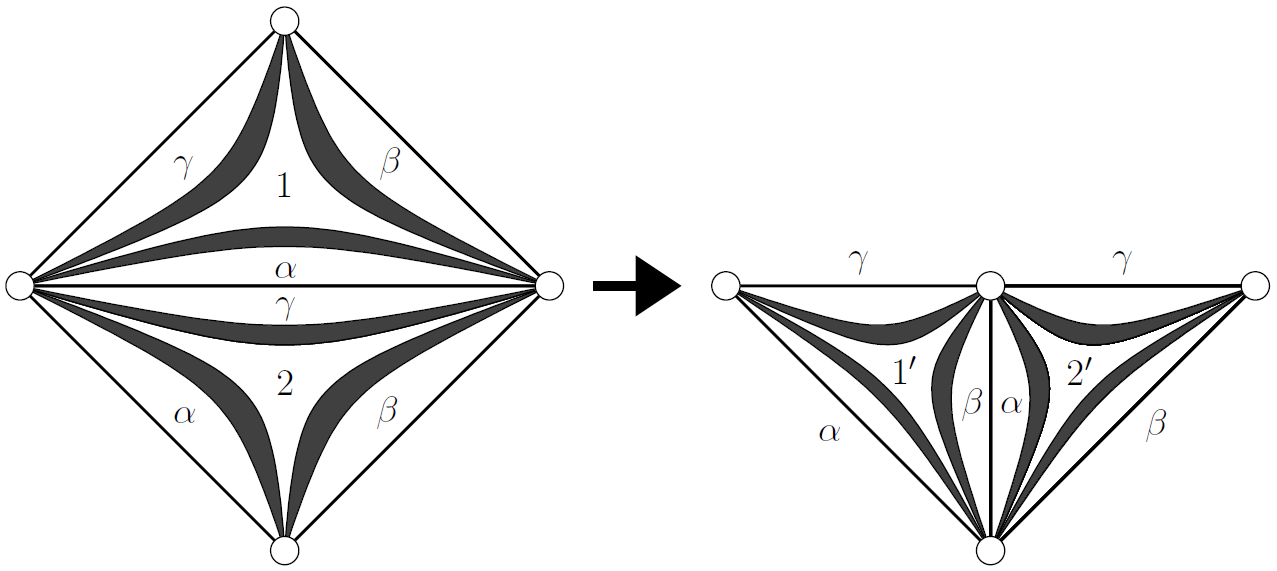}
\caption{\label{transition}Before and after a discrete change in triangulation. $\alpha, \beta, \gamma$ label the three regions of each triangle. Grey regions (bounded by constant-$u$ lines) indicate where the connection is generally non-zero.}
\end{center}
\end{figure}

Dynamics are implicitly determined by the direction of momentum $N^i(q_\pi)$ for each particle via the equation of motion:
\be
\dot{q}_\pi^i(t) = q_\pi^i(0) + t N^i(q_\pi) ,
\ee
and the triangulation evolves accordingly. Non-trivial, discrete transitions occur when a particle meets an edge. Consider triangles $1$ and $2$ depicted in fig. \ref{transition}.
If the particle at the top of the left hand side of the figure moves downward until it reaches the shared edge, then $1$ closes and $2$ splits into $1^\prime$ and $2^\prime$ (shown on the right hand side) preserving the total number of triangles. The fields are defined in the new triangles by specifying the $\SU(2)$ rotation parameters:
\be
&&\hspace{-20pt} \mathbf{P}_{1^\prime_\gamma} = \mathbf{P}_{1_\alpha} + \mathbf{P}_{2_\gamma} + \mathbf{P}_{1_\gamma}, \hspace{0.25in} \mathbf{P}_{1^\prime_\alpha} = \mathbf{P}_{2_\alpha}, \nonumber \\
&&\hspace{-20pt} \mathbf{P}_{2_\gamma^\prime} = \mathbf{P}_{1_\alpha} + \mathbf{P}_{2_\gamma} + \mathbf{P}_{1_\beta}, \hspace{0.25in} \mathbf{P}_{2_\beta^\prime} = \mathbf{P}_{2_\beta}, \\
&&\hspace{-20pt} \mathbf{P}_{1_\beta^\prime} = \mathbf{P}_{2_\alpha^\prime} = 0.  \nonumber
\ee
Notice $h_e=\mathbb{1}$ along the new edge shared by $1^\prime$ and $2^\prime$ so that ${\bm \chi}_{1^\prime} = {\bm \chi}_{2^\prime} = {\bm \chi}_2$. The fields in all other triangles are unaffected by this transition.

\section{Conclusion}

We have described a system of point particles in 2+1 dimensional gravity in terms of evolving triangulations. A gauge choice for $(\mathbf{A}, \mathbf{e})$ was specified by choosing $a_\Delta$ and ${\bm \chi}_\Delta$ in each triangle. The triangulation evolves according to particle dynamics so that particles remain at vertices at all times. The discrete change in triangulation that occurs when a vertex meets an edge is well-defined.

This construction yields spatial geometries that are the two-dimensional analog of those in \cite{FGZ}. Having $(a_\Delta, {\bm \chi}_\Delta)$ in each triangle allows us to immediately write this data in terms of holonomy-flux variables. Moreover, knowing how $(a_\Delta, {\bm \chi}_\Delta)$ behave under a discrete change in triangulation tells us how the holonomies and fluxes will change. This sets the stage for a semiclassical loop gravity description of the system \cite{Z}.

\textit{Acknowledgements:} The author is grateful to Gabor Kunstatter, Laurent Freidel and John Moffat for helpful conversations during the course of this work.

\end{document}